\documentclass[a4paper,fleqn,usenatbib]{mnras}

\usepackage{aecompl}
\usepackage{graphicx}
\usepackage{times}
\usepackage{booktabs}
\usepackage{float}
\usepackage[english]{babel}
\usepackage{color}
\usepackage{amsmath}
\usepackage{chngcntr}

\usepackage[T1]{fontenc}
\usepackage{ae,aecompl}

\usepackage{graphicx}	% Including figure files
\usepackage{amsmath}	% Advanced maths commands
\usepackage{amssymb}	% Extra maths symbols

\title[kSZ$^2$-21 cm correlations]
{Measuring patchy reionisation with kSZ$^2$-21 cm correlations}

\author[Ma et al.]{
Q. Ma$^{1,2,6}$,\thanks{E-mail: maqb@mpa-garching.mpg.de}
K. Helgason$^{2,3}$,
E. Komatsu$^{2,5}$,
B. Ciardi$^{2}$,
A. Ferrara$^{4,5}$
\\
$^1$ Purple Mountain Observatory, Chinese Academy of Sciences, Nanjing 210008, China\\
$^2$ Max-Planck-Institut f\"ur Astrophysik, Karl-Schwarzschild-Stra\ss e 1, D-85748 Garching bei M\"unchen, Germany\\
$^3$ Centre for Astrophysics and Cosmology, University of Iceland, Dunhagi 5, 107 Reykjav\'ik, Iceland \\
$^4$ Scuola Normale Superiore, Piazza dei Cavalieri 7, I-56126 Pisa, Italy \\
$^5$ Kavli Institute for the Physics and Mathematics of the Universe (Kavli IPMU, WPI), Todai Institutes for Advanced Study, \\
University of Tokyo, Kashiwa 277-8583, Japan\\
$^6$ University of Chinese Academy of Sciences, Beijing 100049, China
}

\date{Accepted XXX. Received YYY; in original form ZZZ}

\pubyear{2017}

\begin{document}
\label{firstpage}
\pagerange{\pageref{firstpage}--\pageref{lastpage}}
\maketitle

% Abstract of the paper
\begin{abstract}
We study cross-correlations of the kinetic Sunyaev-Zel'dovich effect
 (kSZ) and 21 cm signals during the epoch of reionisation (EoR) to
 measure the effects of patchy reionisation. Since the kSZ effect is
 proportional to the line-of-sight velocity, the kSZ-21 cm cross
 correlation suffers from cancellation at small angular scales. We thus
 focus on the correlation between the kSZ-squared field (kSZ$^2$) and 21
 cm signals. When the global ionisation fraction is low
 ($x_e\lesssim 0.7$), the kSZ$^2$
 fluctuation is dominated by rare ionised bubbles which leads to an anti-correlation with the 21 cm signal. When $0.8\lesssim x_e<1$, the correlation
 is dominated by small pockets of neutral regions, leading to a positive
 correlation. However, at very high redshifts when $x_e<0.15$, the spin
 temperature fluctuations change the sign of the correlation from
 negative to positive, as weakly ionised regions can have strong 21 cm
 signals in this case. To extract this correlation, we find that Wiener filtering
 is effective in removing large signals from the primary CMB
 anisotropy.
 The expected signal-to-noise ratios for a
 $\sim$10-hour integration of upcoming Square Kilometer Array data cross-correlated with maps from the current generation of CMB observatories with
3.4~$\mu$K arcmin noise and 1.7~arcmin beam
 over 100~deg$^2$ are 51, 60, and 37 for $x_e=0.2$, 0.5, and 0.9, respectively.
\end{abstract}

% Select between one and six entries from the list of approved keywords.
% Don't make up new ones.

%%%%%%%%%%%%%%%%%%%%%%%%%%%%%%%%%%%%%%%%%%%%%%%%%%

%%%%%%%%%%%%%%%%% BODY OF PAPER %%%%%%%%%%%%%%%%%%
%
\section{INTRODUCTION}
\label{sec:intro}
The kinetic Sunyaev--Zel'dovich (kSZ) effect \citep{kSZ1980} is a powerful
probe of the physics of the epoch of reionisation (EoR), as it is sensitive
to patchiness of ionised bubbles \citep[see][and references
therein]{Park2013}. Measurements of the power spectrum of the kSZ,
however, face two challenges. First, we can only measure the sum of the kSZ power spectra
from the EoR and post EoR, and the latter is larger than the
former by at least a factor of two
\citep[e.g.,][]{Shaw2012,Park2016}. Second, the kSZ power spectrum is
sub-dominant compared to other components including the
primary cosmic microwave background (CMB) temperature anisotropy, foreground sources, and the thermal Sunyaev--Zel'dovich effect
\citep{George2015spt}, and thus inaccurate modeling of these components
results in inaccurate estimation of the kSZ power spectrum.

These issues arise because we have no redshift information  of the
kSZ. Cross-correlating the kSZ with 21 cm fluctuations from neutral
hydrogen atoms would allow us to
do ``tomography'' of the kSZ as a function of redshift because the
frequencies of measured 21 cm lines can be translated into $z$
\citep{Alvarez2006,Adshead2008,Alvarez2015}. This cross-correlation not
only helps measurements of the kSZ from the EoR, but also 21
cm signals, as the latter are contaminated by the
Galactic and extragalactic foreground emission and instrumental systematics arising from, e.g.,
miscalibration of gains, polarisation-to-intensity leakages, etc
\citep{Patil2017}, which are not correlated with the CMB
data\footnote{There is still a possibility that unresolved radio
sources have both the 21 cm and CMB data that can correlate.}.

In this paper we use semi-numerical simulations of the EoR
\citep{Mesinger2011} to study the cross-correlation between kSZ and
21 cm signals. In particular, we investigate the
cross-correlation between {\it squared} kSZ fluctuations and 21
cm signals. This was considered in the context of cross-correlation with
weak lensing by the large-scale structure \citep{dore2004}, as well as with galaxies
\citep{Hill2016ksz,Ferraro2016ksz2} in a low redshift universe. This
approach works better because it avoids line-of-sight cancellation of
the kSZ-density correlation.

The rest of the paper is organised as follows.
We describe our simulations in section~\ref{sec:simu}.
In section~\ref{sec:res} we first test the fidelity of our simulated kSZ maps by using the kSZ auto power spectrum, and then show that
the kSZ-21 cm correlation suffers from line-of-sight cancellation on
small angular scales. We then present our results for the kSZ$^2$-21 cm
correlations. In section~\ref{sec:sn} we discuss the detectability of the
kSZ$^2$-21 cm correlations from the EoR by calculating signal-to-noise
ratios of some representative experimental configurations.
We conclude in section~\ref{sec:dis}.

Throughout this paper we use the best-fitting cosmological parameters
from {\it Planck}+WP+highL+BAO data \citep{Planck2013}: cosmological
constant $\Omega_{\Lambda}=0.6914$, matter density $\Omega_{M}=0.3086$,
baryon matter density $\Omega_{b}=0.0483$, scalar spectral index
$n_{s}=0.9611$, matter fluctuation amplitude $\sigma_8=0.8288$, and
Hubble constant $H_0=67.77\,{\rm Km/s/Mpc}$ ($h=0.6777$).
%
%%%%%%%%%%%%%%%%%%%%%%%%%%%%%%%%
\section{SIMULATIONS}
\label{sec:simu}
Temperature anisotropy due to the kSZ effect is given by \citep{kSZ1980}:
\begin{equation}
\label{ksz_eq}
\delta T_{\rm kSZ} (\hat{\gamma})=-{T_0}\int {\rm d} \tau e^{-\tau} \frac{\hat{\gamma} \cdot \pmb{v}}{c},
\end{equation}
where $T_0=2.728\,\rm K$ is the CMB temperature at $z=0$, $c$ the
speed of light, $\pmb{v}$ the bulk peculiar velocity of ionised gas,
$\hat{\gamma}$ a unit vector of the line-of-sight
(LOS), and $\tau$ the optical depth of free electron scattering, i.e.,
\begin{equation}
\label{tau_eq}
{\rm d} \tau = \sigma_{T} N_{b,0} (1+z)^2 (1+\delta) x_{e} {\rm d} s,
\end{equation}
where $\sigma_{T}$ is the Thomson scattering cross-section,
$N_{b,0}=0.2\,(\Omega_b\,h^2/0.022)~{\rm m^{-3}}$ the average atomic number
density at $z=0$, $\delta$ the matter overdensity, $x_{e}$ the
ionisation fraction (assuming the same ionisation fraction for HI and
HeI and no HeII ionised during the EoR), ${\rm d} s = c/H(z) {\rm d} z$ the
differential comoving distance $s$, and
$H(z)=H_{0}\sqrt{\Omega_{M}(1+z)^3+\Omega_{\Lambda}}$ the Hubble
expansion rate at $z$.

As $\tau =\int_{0}^{z} {\rm d} \tau$ is usually much smaller than unity,
we shall ignore fluctuations in $e^{-\tau} \approx 1-\tau$.
Then the kSZ is determined by a specific ionised momentum field defined
by $\pmb{q}\equiv x_{e} (1+\delta) \pmb{v}$.

The offset of the 21 cm brightness temperature from the CMB, $\delta T_{\rm 21cm}$, is expressed as \citep{Mesinger2011}:
\begin{equation}
\label{eq_21cm}
\delta T_{\rm 21cm} = \Psi_{\rm 21cm} (1- x_{e}) (1+\delta) \left(\frac{H}{{\rm d}v_{s}/{\rm d}s+H}\right) \left[ 1-\frac{T_{0}(1+z)} {T_{\rm S}} \right],  %%(\hat{\gamma}, z)
\end{equation}
where $\Psi_{\rm 21cm} \approx 27{\rm mK} [(1+z)/10]^{1/2}$,
$dv_{s}/ds$ is a gradient of the comoving velocity along the LOS, and
$T_{\rm S}$ is the gas spin temperature.
As in this paper we are only considering the 21 cm signals perpendicular to the LOS, we ignore the redshift space distortion term, $dv_{s}/ds$, in the denominator.

\begin{figure*}
\centering
    \includegraphics[width=0.95\linewidth]{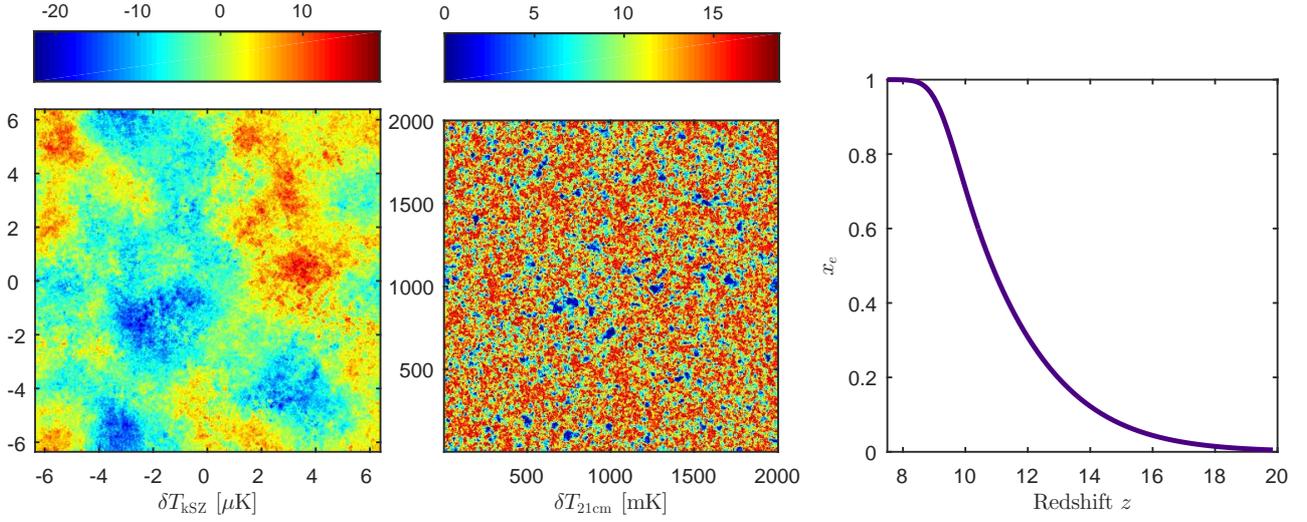}
    \caption{{\bf Left panel:} One realisation of kSZ map ($\delta T_{\rm kSZ}$ in
 units $\rm \mu K$). The size of the map is approximately
 $12.7^\circ\times 12.7^\circ$. {\bf Middle panel:} A slice of 21 cm brightness temperature ($\delta T_{\rm 21cm}$
 in units $\rm mK$) at $x_{e}=0.5$ ($z=10.8$). The physical linear size of
 this map is $2000$ comoving Mpc.
 {\bf Right panel:} Reionisation history ($x_{e}$) as a function of redshift.}
\label{image}
\end{figure*}

We use the semi-numerical simulation code 21cmFAST
\citep{Mesinger2011} to calculate the ionisation fraction, matter
density, and peculiar velocity fields.
The simulations start at $z=30$ with a box of comoving $2000\,\rm
Mpc$ per side and a grid of $400^3$ cells.
This gives a resolution of $5\,\rm Mpc$ per cell.
The volume of the box is large enough to encompass the redshift range
from $z=20$ to $z=7.5$, where we save outputs.
We integrate Eq.~\ref{ksz_eq} in each
resolution element through the box to obtain 2-D kSZ maps (see left panel of Fig.~\ref{image}), by using the
plane-parallel approximation instead of tracing the actual LOS.
21cmFAST also provides the 21 cm field, as shown in the middle panel of Fig.~\ref{image}.

Regarding the ionisation processes, 21cmFAST keeps track of both UV and X-ray radiation, i.e. a cell is ionised if $\zeta_{\rm UV}f_{\rm coll}(R)\ge 1-x_{e,\rm X}$,
where $f_{\rm coll}$ is the collapsed fraction inside a sphere with a
radius $R$, $\zeta_{\rm UV}$ is the ionising efficiency factor of UV photons, and $x_{e,\rm X}$ is the fraction ionised by X-ray radiation \citep[for details, see ][]{Mesinger2013}.
The X-ray emission is due to stellar remnants, e.g. X-ray binaries whose
ionising efficiency factor is related to the star formation,
$\zeta_{\rm X}=({N_{\rm X}}/10^{56}~{\rm M}_{\odot})({f_{\ast}}/{0.1})$,
where  $N_{\rm X}$ is the X-ray  ($\ge 0.3\,{\rm KeV}$) photon
number emitted per solar mass during the whole life of stars (in units of ${\rm M}_{\odot}^{-1}$), and $f_{\ast}$ is the fraction of collapsed baryons converted into stars \citep{Mesinger2013}.
The values we adopted for these parameters are $\zeta_{\rm UV}=31.5$ and $\zeta_{\rm X}=1$.

We generate 20 realisations to obtain good statistics. Each realisation
gives three independent 2-D kSZ maps along three axes of the snapshot;
thus, we have 60 realisations of kSZ maps and the corresponding 21 cm
fields.

The right panel of Fig.~\ref{image} shows the reionisation
history of our simulations.
The reionisation completes at $z\approx 8$, with
half-ionisation at $z=10.8$. Assuming that hydrogen is fully ionised at
$z<8$ and helium is singly ionised ($x_{\rm HeII}=1$) at $z>3$ and
doubly ionised ($x_{\rm HeIII}=1$) at $z<3$, we obtain an optical depth
of $\tau \approx 0.0996$, which is high compared to the latest
determination by Planck \citep{Planck2016}.

The left panel of Fig.~\ref{image} shows one realisation of the kSZ map,
while the middle panel shows a slice of the 21 cm brightness temperature at
$z=10.8$ ($x_{e} = 0.5$). The kSZ signal is primarily generated by a
long-wavelength peculiar velocity field modulated by small-scale
electron density fluctuations \citep{Hu2000}. This is the reason that
the kSZ map is dominated visually by long-wavelength modes.

%
%
%%%%%%%%%%%%%%%%%%%%%%%%%%%%%%%%%%%
\section{Angular Power Spectra}
\label{sec:res}
In the flat-sky approximation, the angular spectrum is computed by
\begin{equation}
<\tilde{X}^{\ast}(\pmb{l})\tilde{Z}(\pmb{l'})> =  C_{X-Z}(l)\delta_{\rm D}(\pmb{l}-\pmb{l'}),
\end{equation}
where $\delta_{\rm D}$ is the Dirac delta function, $X$ and $Z$ are maps, and $\tilde{X}$ and $\tilde{Z}$ are their Fourier transforms:
\begin{equation}
\tilde{X}(\pmb{l})=\frac{1}{2\pi}\int d^2\hat{\gamma} X(\hat{\gamma}) e^{-i\pmb{l}\cdot \hat{\gamma}}.
\end{equation}
If $X=Z$, $C_{X-X}(l)$ is the auto power spectrum of a 2-D map $X$.
%%%%%%%%%%%%%%%%%%%%%%%%%%%%%%%%%%%
\subsection{kSZ auto power spectrum}
\begin{figure}
\centering
\includegraphics[width=0.88\linewidth]{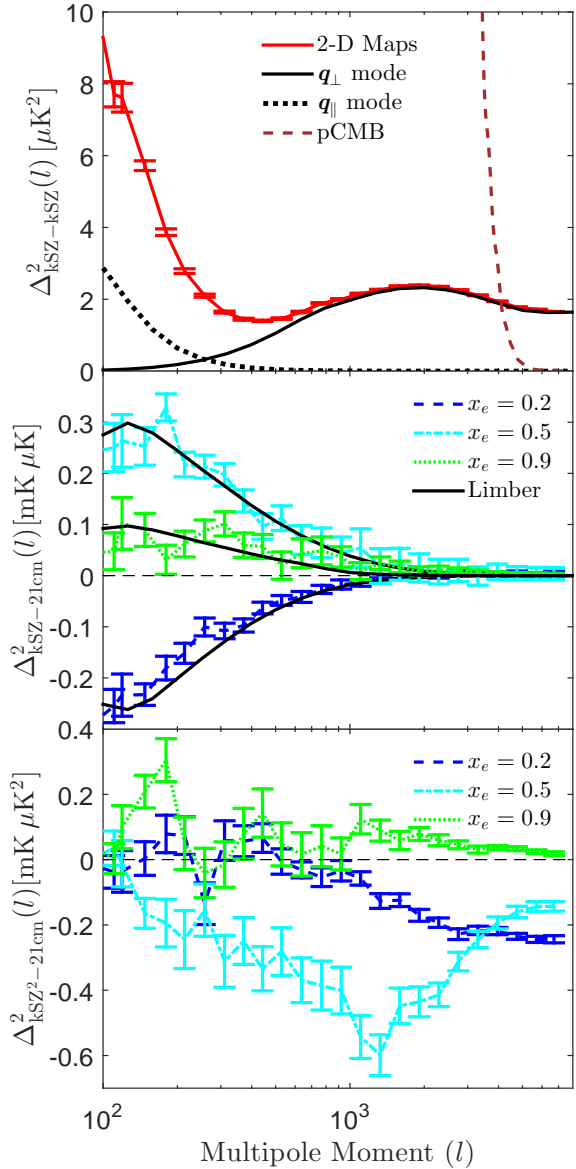}
\caption{{\bf Top panel:} kSZ auto power spectrum, $\Delta^{2}(l)=l(l+1)C(l)/2\pi$, averaged over 60 realisations (solid
 red line). The error bars are the errors on the average (i.e.,
 r.m.s. scatter divided by $\sqrt{60}$). The solid (dotted) black line
 shows the power spectrum of transverse (longitudinal) momentum derived
 from the 3-D power spectra integrated over redshifts using Limber's
 approximation. The dashed line shows the primary CMB power spectrum.
 {\bf Middle panel:} kSZ-21 cm cross power spectra at $x_{e}=0.2$ ($z=13$;
 dashed blue), $x_{e}=0.5$ ($z=10.8$; dash-dotted cyan), and $x_{e}=0.9$
 ($z=9.3$; dotted green). The solid black lines are the corresponding
 Limber results. The horizontal dashed line shows zero.
 {\bf Bottom panel:} Same as the middle panel but for kSZ$^2$--21 cm cross correlations.}
\label{pscl_all}
\end{figure}
In the top panel of Fig.~\ref{pscl_all}, we show the kSZ auto power
spectrum with the error bars on the average derived from 60 realisations.

To check the fidelity of the maps, we also compute the kSZ power spectrum by
directly integrating the 3-D power spectra of specific momentum
fields over redshifts. The momentum fields consist of two components: the transverse
mode whose direction is perpendicular to the wavenumber, $\pmb{k}$,
i.e., $\pmb{k}\perp{\pmb{q}_{\perp}}$, and the longitudinal mode,
$\pmb{k}\parallel {\pmb{q}_{\parallel}}$. Using Limber's approximation
\citep{Limber1953}, we obtain \citep{Park2013,Alvarez2015}
\begin{eqnarray}
\label{q_trans}
C_{\pmb{q}_{\perp}}(l)&=&\left(\frac{\sigma_{T}N_{b,0}T_{0}}{c}\right)^2\int (1+z)^{4} \frac{{\rm d} s}{s^2}~e^{-2\tau} \frac{P_{\pmb{q}_{\perp}}(\frac{l}{s},z)}{2},\\
% \nonumber
\label{q_para}
C_{\pmb{q}_{||}}(l) &=&
\frac1{l^2}\int {\rm d} s~ \Psi_{||}^2 \frac{P_{\delta \delta}(\frac{l}{s},z)}{(l/s)^2},
\end{eqnarray}
where in the last line we have used linear theory to relate the
longitudinal velocity with the matter density. Here, $P_{\delta \delta}$ is the power spectrum of matter
density,  $\Psi_{||}\equiv \frac{T_{0}}{c D} {\rm d} (a \dot{D}
\frac{{\rm d} \tau}{{\rm d} s} )/{\rm d}s$, $D$ is the growth factor of
linear matter density fluctuations, and $\dot{D}$ is the time derivative
of $D(z)$. We evaluate these integrals using $P_{\pmb{q}_{\perp}}$ and
$P_{\delta \delta}$ measured from the simulations.

The factor of $l^{-2}$ in Eq.~\ref{q_para} is a consequence of LOS
cancellation; namely, the kSZ is caused by the LOS component of
velocities, and the longitudinal velocities are parallel to
$\pmb{k}$. As the longitudinal modes change signs along the LOS, short
wavelength modes suffer from cancellations
\citep{Vishniac1987}. Therefore, the longitudinal modes dominate at large
angular scales.

We find that the Limber formula agrees well with
the kSZ power spectrum measured from the maps at $l\gtrsim 10^3$. However,
at much lower multipoles, the power measured from the maps is substantially
larger than the Limber formula for the longitudinal mode. This
large-scale mismatch originates from boundary effects as the LOS
cancellation does not occur at the near/far boundaries of our lightcone
(i.e. at $z\sim7.5$ and $z\sim20$).

\subsection{kSZ-21 cm correlation}
The middle panel of Fig.~\ref{pscl_all} shows the cross power spectra of
kSZ with 21 cm at $x_{e}=0.2$, 0.5 and 0.9, together with the results
from the Limber formula \citep{Alvarez2006}\footnote{We have added a term including spin temperature $T_{\rm S}$, while when $T_{\rm S} \gg T_{\rm CMB}$, Eq.~\ref{eq_21cm_ksz} can be simplified to the Eq.~17 in \cite{Alvarez2006}.}:
\begin{align}
\label{eq_21cm_ksz}
C_{\rm kSZ-21cm}(l) &= \frac1{l^2} \Psi_{||} \Psi_{21cm} \kappa \Big\{- x_{e}P_{\delta \delta_{x}}\left(\frac{l}{s},z\right) \nonumber \\
& \qquad{} + (1-x_{e})\left[P_{\delta \delta}\left(\frac{l}{s},z\right)+P_{\delta \delta_{\kappa}}\left(\frac{l}{s},z\right)\right]\Big\},
\end{align}
where $\kappa \equiv 1-T_{0}(1+z)/T_{\rm S}$, $P_{\delta  \delta_{x}}$
is the cross power spectrum of matter density and ionisation fraction
fluctuations, and $P_{\delta  \delta_{\kappa}}$ is the cross power
spectrum of matter density and $\kappa$ fluctuations.

Here, the factor $l^{-2}$ is again due to LOS cancellation, as the correlation is
dominated by the longitudinal modes correlated with density fields. The
correlation between the transverse modes and density fields involves a
three-point correlation of $(\pmb{v}\delta)_\perp\delta$, which vanishes
for Gaussian fluctuations.
We evaluate Eq.~\ref{eq_21cm_ksz} using all the 3-D power spectra
measured from the simulations. We find excellent agreement between the
power spectra measured from maps and the Limber results, which indicates
that the contribution from the transverse modes is indeed negligible.

\begin{figure}
    \centering
\includegraphics[width=0.95\linewidth]{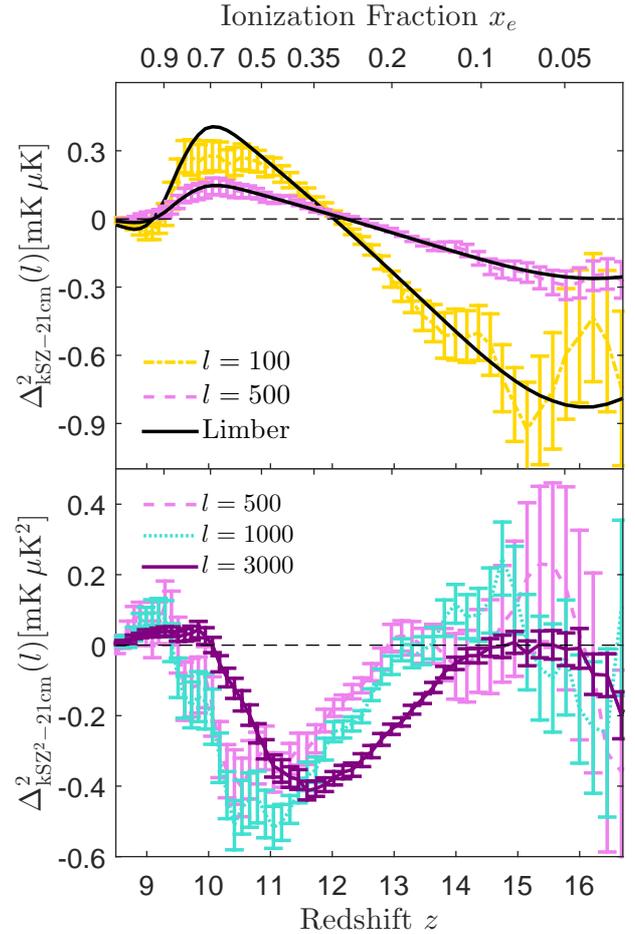}
\caption{{\bf Top panel:} Redshift evolution of the kSZ-21 cm cross power spectra at
 $l=100$ (dash-dotted yellow) and 500 (dashed violet).
The black lines are the corresponding Limber results.
 The horizontal dashed line shows zero.
 {\bf Bottom panel:} Same as the top panel but for kSZ$^2$--21 cm cross correlations of $l=500$, 1000, and 3000.}
\label{ps_ksz_vz}
\end{figure}

In the top panel of Fig.~\ref{ps_ksz_vz} we show the redshift evolution of the kSZ-21 cm
cross power spectrum at $l=100$ and 500 together with the Limber
results. Because the correlations are overestimated at the lightcone boundaries, we crop the evolution to $z=16.7-8.5$.
Our results are in agreement with \cite{Alvarez2006}, apart from
high redshifts where the spin temperature makes the cross-correlation
more negative. This happens because $T_{S}$ strongly correlates with the matter density,
i.e., $P_{\delta\delta_{\kappa}} \gg P_{\delta\delta}$, the magnitude of
the kSZ-21 cm correlation becomes larger as $z$ increases.
%

%
%%%%%%%%%%%%%%%%%%%%%%%%%%%%%%%%%%%%%%%
\subsection{kSZ$^2$-21 cm correlation}
In order to detect the cross-correlation between kSZ and 21 cm signals, we need to overcome the LOS cancellation. One way to achieve this is to cross-correlate {\it
squared} kSZ fields with density fields \citep{dore2004}. Then the correlation between kSZ$^2$ and 21 cm signals would persist at small angular scales.

To avoid the contamination of boundary effects at large scales, we
remove the kSZ signals at $l<100$ before squaring, and only focus on the correlations at $z=16.7-8.5$ and $l>100$.

In the bottom panel of Fig.~\ref{pscl_all} we show the kSZ$^2$-21 cm cross power spectra at
$x_{e}=0.2$, 0.5 and 0.9 as a function of multipoles with the error bars
on the average of 60 realisations, while in the bottom panel of Fig.~\ref{ps_ksz_vz}
we show redshift evolution of the spectra at $l=500$, 1000, and
3000. These spectra evolve rapidly with ionisation fraction. For
example, at $x_{e}=0.2$ ($z=13$) kSZ$^2$ correlates negatively with
the 21 cm signal at small scales ($l>10^3$), but no significant
correlations are visible at large scales ($l<10^3$), whereas at $x_{e}=0.5$ ($z=10.8$) the
kSZ$^2$-21 cm correlation is negative in the entire multipole
range. At a later stage of reionisation, when $x_{e}=0.9$ ($z=9.3)$,
the correlation turns slightly positive.

\begin{figure}
	\centering
	\includegraphics[width=0.95\linewidth]{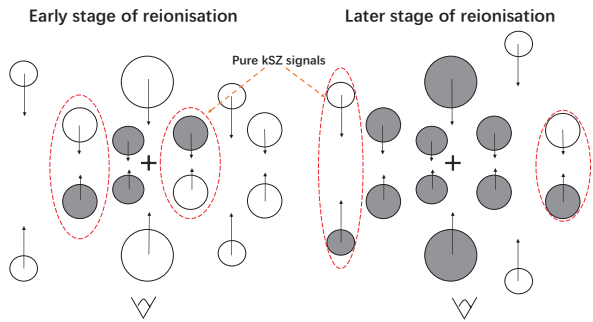}
 \caption{Simplified diagram to illustrate the dependence of kSZ$^2$ fluctuations on the reionisation phase, assuming the velocities along LOS appear as pairs. During the early stages of reionisation (left), when the Universe is lowly ionised, the ionised bubbles (gray) have a larger probability to meet neutral pair partners (white) than ionised ones and then remain pure kSZ signals (embedded in red dashed lines). In this case, the kSZ$^2$ fluctuations are dominated by the ionised bubbles. At a later stage of reionisation (right), when the Universe is highly ionised, the situation is reversed and the neutral regions have a higher probability to remain pure kSZ signals compared to the ionised bubbles. The kSZ$^2$ fluctuations are thus dominated by the neutral regions.}
\label{ksz_car}
\end{figure}

The evolution of kSZ$^2$-21 cm cross power spectra can be understood in terms of the highly
non-Gaussian nature of reionisation. When ionisation is low, e.g. at $x_{e}<0.7$ ($z>10$), the
kSZ$^2$-21 correlation is dominated by rare ionised bubbles (see the left part of illustration in Fig.~\ref{ksz_car}) that
anti-correlate with the 21 cm signal arising from the neutral medium,
resulting in negative cross spectra.

On the other hand, kSZ$^2$ shows positive correlations with 21 cm at $x_{e}<0.15$ ($z>14$), when the spin temperature dominates the 21 cm signal. This is because $T_S$ correlates strongly with the matter density at these epochs.

When the Universe is highly ionised, e.g. at $x_{e}>0.8$ ($z<9.7$), kSZ$^2$ fluctuations are dominated by the small remains of the neutral medium (see the right part of illustration in Fig.~\ref{ksz_car}),
resulting in positive cross spectra with 21 cm signals.
%
%%%%%%%%%%%%%%%%%%%%%%%%%%%%%%%%%%%
\section{Signal-to-noise Ratio}
\label{sec:sn}
We calculate the expected signal-to-noise ratios (S/N) of the kSZ$^2$-21 cm
correlations. We adopt specifications
similar to those of LOFAR \citep{Vrbanec2016} and SKA \citep{Koopmans2015} for 21
cm observations, and those of the current generation of ground-based CMB
observatories such as SPT-3G \citep{SPT3G} and Advanced ACT
\citep{AdvACT}.

We estimate the S/N per multipole bin as \citep{dore2004}
\begin{equation}
\left( \frac{S}{N} \right)^2 = \frac{f_{\rm sky}(2l+1) l_{\rm bin} C_{\rm kSZ^2, 21cm}^{2}} {C_{\rm CMB^2} (C_{\rm 21cm} + N_{\rm 21cm})+C_{\rm kSZ^2,21cm}^{2}},
\label{eq:sn}
\end{equation}
where $l_{\rm bin}\approx 0.46l$ ($({\rm log_{10}} l)_{\rm bin}=0.2$) is
the bin width at a given $l$, $f_{\rm sky}$ the fraction of sky observed
by both CMB and 21 cm experiments, $C_{\rm CMB^2}$ the power spectrum of
CMB-squared, and $C_{\rm 21cm}$ and $N_{\rm 21cm}$ are the auto spectrum
of the 21 cm signal and its noise, respectively.

In the CMB maps we include primary CMB and the thermal SZ effect
\citep{Dolag2015}. To this we also add the Poisson and clustered power from dusty
star-forming galaxies and the Poisson power from radio galaxies, for which we
use the values estimated by the SPT Collaboration \citep{George2015spt}.
Finally we add Gaussian, white instrumental noise of $3.4~\mu{\rm
K}$~arcmin which corresponds to a noise
per pixel of $\sigma_{\rm pix}=2 \,\mu{\rm K}$ with our pixel size of 1.7 arcmin. We then use these maps
to calculate $C_{\rm CMB^2}$ in Eq.~\ref{eq:sn}.

The auto spectra of 21 cm signals, $C_{\rm 21cm}$, come from our simulations, and the
noise power is given by $N_{\rm 21cm}=[(1+z)/9.5]^2\sigma_{\rm pix}^2\theta_{\rm
FWHM}^2$ \citep{dore2004}.
We adopt $\theta_{\rm FWHM}=3.5\,\rm arcmin$ and  $\sigma_{\rm
pix}=76\,\rm mK$ at 150MHz ($z\approx8.5$) for LOFAR
\citep{Vrbanec2016}, assuming 600 hours of integration and a bandwidth
of 0.5~MHz. SKA will have a superior angular resolution of $\theta_{\rm
FWHM}=1\,\rm arcmin$ \citep{Koopmans2015}, and a lower noise level.
We adopt $\sigma_{\rm pix}=10\,\rm mK$ at 150~MHz, which
corresponds to $\sim$10 hours of integration and 1~MHz of bandwidth.

Finally, we assume that both CMB and 21 cm experiments will have an
overlapping region of 100~deg$^2$ ($f_{\rm sky}=0.0024$) for SKA, and
25~deg$^2$ for LOFAR. These parameters are listed in Table~\ref{tab1}.

Note that both squared fields and 21~cm signals are non-Gaussian, but
Eq.~\ref{eq:sn} is valid only for Gaussian fields. Thus, the S/N
estimate given here is only approximate.

\begin{table}
\label{tab1}
\centering
\begin{tabular}{c c c c}
\hline\hline
Experiment & $\theta_{\rm FWHM}$ &  $\sigma_{\rm pix} (\mu \rm K)$ & $f_{\rm sky}$\\
\hline
CMB    &    1.7             &      2                          &   0.0024 \\
LOFAR      &    3.5             &      $76\times10^3$             &   0.0006 \\
SKA        &    1.0             &      $10\times10^3$             &   0.0024 \\
\hline
\end{tabular}
\caption{Characteristics of CMB and 21 cm experiments.}
\end{table}

We find that the S/N is small  ($\ll 1$) for any combinations of CMB and 21 cm
experiments, mainly because of the large noise from the primary CMB signal. To mitigate this problem, in the next section we apply the commonly adopted Wiener filtering \citep{dore2004}.

\subsection{Wiener filtering}
To suppress primary CMB ``noise'', we apply the following filter \citep{dore2004, Hill2016ksz, Ferraro2016ksz2}
\begin{equation}
\label{eq:wf}
F(l)=\frac{C_{\rm kSZ}(l)} {C_{\rm kSZ}(l) + C_{\rm pCMB}(l) + C_{\rm fore}(l)},
\end{equation}
where $C_{\rm pCMB}$ is the primary CMB power spectrum, and $C_{\rm
fore}$ is the sum of the foreground terms including the thermal SZ,
dusty star-forming galaxies and radio galaxies.
As Wiener filtering will automatically suppress low $l$ power, we do not crop the CMB fluctuations at $l<100$ before applying the filtering.

\begin{figure}
    \centering
    \includegraphics[width=0.95\linewidth]{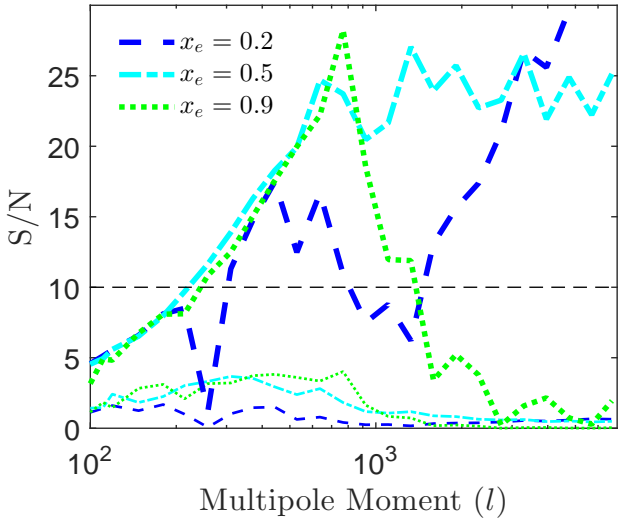}
    \caption{Predicted S/N of the kSZ$^2$-21 cm correlations per
 multipole bin after Wiener
 filtering for $x_{e}=0.2$ ($z=13$; dashed blue), $x_{e}=0.5$ ($z=10.8$; dash-dotted cyan), and $x_{e}=0.9$ ($z=9.3$; dotted green).
    The thick and thin lines refer to the SKA
 and LOFAR case, respectively. The horizontal line marks S/N $=10$.}
\label{sn_ratio_vl}
\end{figure}
\begin{figure}
    \centering
    \includegraphics[width=0.95\linewidth]{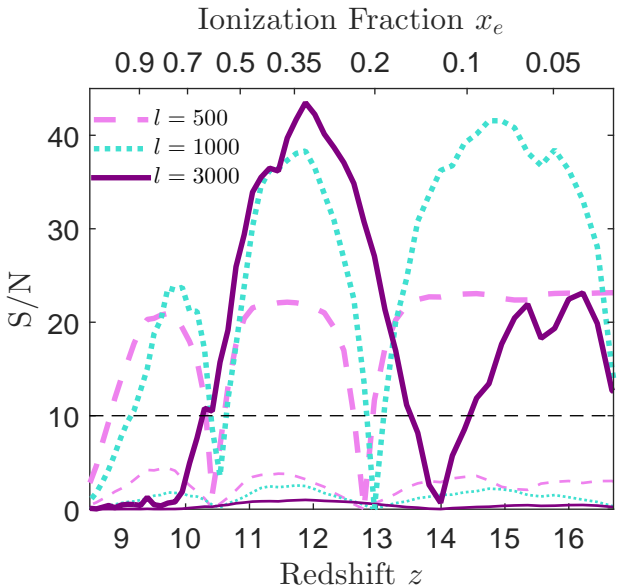}
    \caption{Predicted S/N of the kSZ$^2$-21 cm correlations after
 Wiener filtering as a function of redshift for multipole bins of $l=500$ (dashed violet),
 1000 (dotted turquoise), and 3000 (solid purple).
     The thick and thin lines refer to the SKA
 and LOFAR case, respectively. The horizontal line shows S/N $=10$.}
\label{sn_ratio_vz}
\end{figure}
In Fig.~\ref{sn_ratio_vl} we show the predicted S/N as a function of
multipoles at $x_{e}=0.2$, 0.5 and 0.9 after filtering.
We find that the S/N for LOFAR is always below 5. SKA though is much more promising, having
a S/N per multiple bin $>10$ at $x_e=0.9$ in the range $l=200-1000$,
at $x_{e}=0.5$ for $l>200$, and at $x_{e}=0.2$ for $l\sim400$ and $l>2000$.

In Fig.~\ref{sn_ratio_vz} we show the S/N as a function of redshift for
$l=500$, 1000 and 3000. SKA is expected to have a S/N $>10$ over a wide redshift
range, except two narrow gaps at $z\sim 10.5$ ($x_{e}\sim 0.6$) and
$z\sim 13$ ($x_{e}\sim 0.2$), where the correlations are very weak (see the bottom panel of Fig.~\ref{ps_ksz_vz}).

\begin{figure}
    \centering
    \includegraphics[width=0.95\linewidth]{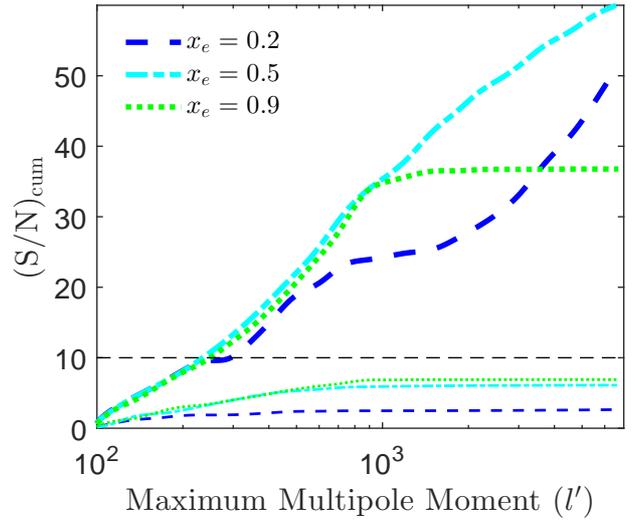}
 \caption{Cumulative S/N as a function of the maximum multipole.
The meaning of the lines is the same as that in Fig.~\ref{sn_ratio_vl}.}
\label{cum_sn_ratio_vl}
\end{figure}
To obtain the total S/N rather than the S/N per multipole bin, we compute
the cumulative S/N as
\begin{equation}
\left(\frac{S}{N} \right)^2_{\rm cum} (<l')=
 \displaystyle\sum_{i}
\frac{f_{\rm sky} (2l_{i}+1) \Delta l_{i}  C_{\rm kSZ^2, 21cm}^{2}} {C_{\rm CMB^2} (C_{\rm 21cm} + N_{\rm 21cm})+C_{\rm kSZ^2,21cm}^{2}},
\end{equation}
where $i$ denotes the $i$th bin, and $l_i$ is the central multipole of the $i$th
bin with bin width $\Delta l_{i}$. We display it in Fig. ~\ref{cum_sn_ratio_vl} as a function of the maximum multipole $l'$.
The cumulative S/N for LOFAR is $<10$, while
the SKA would have S/N as large as 51.4, 60.2, and 36.8 at $x_{e}=0.2$,
0.5 and 0.9, respectively.

\section{Conclusions and Discussion}
\label{sec:dis}
We have used the semi-numerical 21cmFAST simulations
\citep{Mesinger2011} to study cross-correlations of the kSZ and 21 cm
signals from the EoR. As the kSZ-21 cm correlation suffers from the
line-of-sight cancellation at small angular scales, we have focused on
the cross-correlation between {\it squared} kSZ fields and 21 cm signals. We find that the line-of-sight cancellation is mitigated
and the cross-correlation persists at small angular scales.

The goal of this work is to capture the general behavior of the kSZ$^2$-21 cm cross-correlation signal. We caution that the simplified scheme of 21cmFAST may not be adequate to quantify the signal in detail, and one may need more sophisticated
reionisation simulations \citep[e.g. ][]{Park2013}.
It is also important to acknowledge that we have performed our calculations only for one specific set of cosmological parameters and
reionisation history; thus, we have not investigated how the predicted
signals change when we change the physics of reionisation.
In particular, the optical depth from the latest Planck data
\citep{Planck2016} ($\tau = 0.058\pm 0.012$) is lower than that from our
simulation ($\tau \approx 0.1$), and its impact is not obvious.
Finally, we neglected any correlation between foregrounds in 21 cm and kSZ maps which, if present, would contaminate the reionisation signal. The contribution from foregrounds, such as continuum radio sources, would merit further study as well.

The kSZ$^2$-21 cm cross-correlation signal exhibits interesting features, such as a
sign change according to phases of reionisation. For example, the correlation is positive for ionisation fraction $x_{e}<0.15$ (when density fluctuations dominate the 21 cm signal) and at $x_{e}>0.8$ (when the Universe is highly ionised), while it is negative for intermediate ionisation fractions. Thus, not only is the
cross-correlation powerful for minimising the foreground emission and
instrumental systematics of either kSZ or
21 cm data, but also it offers a powerful probe of
reionisation.

In general, prospects for measuring the kSZ$^2$-21 cm cross-correlation signal are good: SKA cross-correlated with
on-going CMB experiments such as SPT-3G and Advanced ACT should
yield high signal-to-noise ratio measurements of the cross-correlation
over a wide range of multipoles and redshifts.

%%%%%%%%%%%%%%%%%%%%%%%%%%%%%%%%%%%%%
\section*{Acknowledgments}
We acknowledge the helpful discussions with V. Jelic and M. Alvarez.
The tools for bibliographic research are offered by the NASA Astrophysics Data Systems and by the JSTOR archive.
QM is supported by the National Natural Science Foundation of China (Grant Nos 11373068 and 11322328), the National Basic Research
Program (973 Program) of China (Grant Nos 2014CB845800 and
2013CB834900) and the Strategic Priority Research Program The Emergence
of Cosmological Structures (Grant No. XDB09000000) of the Chinese
Academy of Sciences.
EK is supported in part by JSPS KAKENHI Grant Number JP15H05896.
KH acknowledges support from the Icelandic Research Fund, Grant Number 173728-051.

\bibliographystyle{mnras}
\bibliography{ref}
% Don't change these lines
\bsp	% typesetting comment
\label{lastpage}
\end{document}